\documentclass[conference,10pt]{IEEEtran}
\IEEEoverridecommandlockouts
\usepackage{amsfonts}
\usepackage{amsmath}
\usepackage{amssymb} 
\usepackage{array}
\usepackage{bm} 
\usepackage{bbm}
\usepackage{color} 
\usepackage{cases} 
\usepackage{cite}
\usepackage{dsfont}
\usepackage[bookmarks=false]{hyperref}
\hypersetup{
	colorlinks=true,
	linkcolor=blue,
	filecolor=blue,
	citecolor = blue,      
	urlcolor=cyan,}
\usepackage[pdftex]{graphicx}
\DeclareGraphicsExtensions{.eps,.pdf,.png,.jpg,.gif,.jpeg,.pstex}
\usepackage{hhline}
\usepackage{multirow}
\usepackage{makecell}
\usepackage{stfloats}
\usepackage{textcomp}
\usepackage{url}
\usepackage{units} 
\usepackage{verbatim}

\usepackage[caption=false, font=footnotesize]{subfig}
\usepackage[printonlyused]{acronym}
\usepackage[table]{xcolor}

\usepackage{algorithm}
\usepackage{algpseudocode}
\usepackage{tikz} 
\usepackage[utf8]{inputenc}
\usepackage{pgfplots} 
\pgfplotsset{width=10cm,compat=1.9}
\usepackage{pgfplotstable}
\usepackage{xurl} 
\usepackage[colorinlistoftodos,bordercolor=orange,backgroundcolor=orange!20,linecolor=orange,textsize=scriptsize]{todonotes}
\usepackage[T1]{fontenc}
\usepackage{newtxtext,newtxmath}

\algtext*{EndWhile}
\algtext*{EndIf}
\algtext*{EndFor}
\usetikzlibrary {arrows.meta}
\makeatletter

\makeatletter
\algnewcommand{\LineComment}[1]{\Statex \hskip\ALG@thistlm \(\triangleright\) #1}
\makeatother

\newacro{ris} [RIS] {reconfigurable intelligent surfaces}
\newacro{ummimo} [UM-MIMO] {ultra-massive multiple-input multiple-output}
\newacro{irs} [IRS] {intelligent reflecting surfaces}
\newacro{elaa}[ELAA]{extremely large antenna arrays}
\newacro{xlmimo}[XL-MIMO]{extremely large-scale MIMO}
\newacro{thz}[THz]{terahertz}
\newacro{mmwave}[mmWave]{millimeter-wave}
\newacro{aosa}[AoSA]{array-of-sub-arrays}
\newacro{daosa}[DAoSA]{dynamic AoSA}
\newacro{gosa}[GoSA]{group-of-sub-arrays}
\newacro{hmimo}[H-MIMO]{holographic MIMO}
\newacro{dma}[DMA]{dynamic metasurface antennas}
\newacro{capmimo}[CAP-MIMO]{continuous aperture MIMO}
\newacro{fas}[FAS]{fluid antenna systems}
\newacro{sim}[SIM]{stacked intelligent metasurfaces}
\newacro{aes}[AEs]{array elements}
\newacro{ae}[AE]{array element}
\newacro{pwm}[PWM]{planar wave model}
\newacro{swm}[SWM]{spherical wave model}
\newacro{hspwm}[HSPWM]{hybrid
spherical planar wave model}
\newacro{sa}[SA]{sub-array}
\newacro{sas}[SAs]{sub-arrays}
\newacro{arvs}[ARVs]{array response vectors}
\newacro{cs}[CS]{compressed sensing}
\newacro{em}[EM]{electromagnetic}
 
\newcommand{\nth}[1]{{#1}{\text{th}}}
\newcommand{\mbf}[1]{\mathbf{#1}}

\newcommand{\Hpow}{{\sf H}}
\newcommand{\Tpow}{{\sf T}}

\newcommand{\abs}[1]{\left|{#1}\right|}
\newcommand{\norm}[1]{\left\|{#1}\right\|}
\newcommand{\vect}[1]{\mathrm{vec}\left(#1\right)}

\newcommand{\txidx}{\mathrm{T}}
\newcommand{\rxidx}{\mathrm{R}}

\newcommand{\quantidx}{\mathrm{quant}}

\newcommand{\FFsupsc}{\mathrm{FF}}
\newcommand{\SWMsupsc}{\mathrm{SWM}}

\newcommand{\PWMsupsc}{\mathrm{PWM}}

\long\def\comment#1{}

\newfont{\bbb}{msbm10 scaled 700}

\newfont{\bb}{msbm10 scaled 1100}
\newcommand{\CC}{\mbox{\bb C}}

\makeatother

\makeatletter
\def\ps@IEEEtitlepagestyle{%
  \def\@oddfoot{\mycopyrightnotice}%
  \def\@oddhead{\hbox{}\@IEEEheaderstyle\leftmark\hfil\thepage}\relax
  \def\@evenhead{\@IEEEheaderstyle\thepage\hfil\leftmark\hbox{}}\relax
  \def\@evenfoot{}%
}

\def\mycopyrightnotice{%
  \begin{minipage}{\textwidth}
  \centering \scriptsize
    Copyright~\copyright~2026 IEEE. Personal use of this material is permitted. Permission from IEEE must be obtained for all other uses, in any current or future media, including reprinting/republishing this material for advertising or promotional purposes, creating new collective works, for resale or redistribution to servers or lists, or reuse of any copyrighted component of this work in other works.
    
    Accepted for publication in the IEEE NextGCom 2026.
  \end{minipage}
}
\makeatother

\begin{document}

\bstctlcite{IEEEexample:BSTcontrol}

\title{Learning-based near- versus far-field boundaries for ultra-massive MIMO communications\\
\thanks{The work of S. Tarboush was supported by the European Union, through the Horizon Europe Marie Skłodowska-Curie Doctoral Networks Programme “Intelligent sensing and communication as training network for perceptive mobile networks in 6G (ISAC-NEWTON)” under Grant 101169496. The work of G. Caire was supported by the BMFTR Germany in the program of “Souverän. Digital. Vernetzt.” Joint Project 6G-RIC (Project IDs 16KISK030). The work of N.~Kouzayha and T.~Y.~Al-Naffouri was supported by the KAUST Office of Sponsored Research (OSR) under Award No. ORFS-CRG12-2024-6478. The work of H. Sarieddeen was supported by the AUB University Research Board (URB) and Vertically Integrated Projects (VIP) program.
}
}

\author{
    \IEEEauthorblockN{
        Simon Tarboush\IEEEauthorrefmark{1},
        Nour Kouzayha\IEEEauthorrefmark{2},
        Hadi~Sarieddeen\IEEEauthorrefmark{3},
        Tareq~Y.~Al-Naffouri\IEEEauthorrefmark{2},
		and Giuseppe Caire\IEEEauthorrefmark{1}
        }
	\IEEEauthorblockA{
        \IEEEauthorrefmark{1}\small Communications and Information Theory Chair, Faculty of Electrical Engineering and Computer Science,\\ Technische Universit{\"a}t Berlin, 10587 Berlin, Germany.
        }
	\IEEEauthorblockA{
        \IEEEauthorrefmark{2}\small Electrical and Computer Engineering Program, Division of Computer, Electrical and Mathematical Sciences and Engineering (CEMSE),\\ King Abdullah University of Science and Technology (KAUST), Thuwal, 23955-6900, Kingdom of Saudi Arabia.
        }
    \IEEEauthorblockA{
        \IEEEauthorrefmark{3}\small Electrical and Computer Engineering Department, American University of Beirut (AUB), Lebanon.
        \\
        Emails: (simon.tarboush; caire@tu-berlin.de), (nour.kouzayha; tareq.alnaffouri@kaust.edu.sa), (hadi.sarieddeen@aub.edu.lb).
        }
}

\maketitle

\begin{abstract}
Signal processing techniques for wireless communications and sensing fundamentally differ between near-field and far-field propagation regimes. Accurately identifying the applicable propagation region is therefore essential for enabling efficient beamforming and channel estimation in ultra-massive MIMO (UM-MIMO) systems. This paper proposes a fully unsupervised learning framework to distinguish near-field from far-field propagation based solely on received signal measurements, before estimating the communication distance, and without relying on channel state information. The proposed approach exploits spatial signal power variations across subarrays of a UM array as a physics-inspired feature extraction stage, followed by the OPTICS clustering algorithm to infer the communication region. Simulation results under various system configurations and signal-to-noise ratio (SNR) levels demonstrate that the proposed method accurately identifies the near-field and far-field regions, showing agreement with theoretical boundaries.
\end{abstract}

\begin{IEEEkeywords}
Near-/far-field classification, planar wave propagation, spherical wave propagation, unsupervised learning.
\end{IEEEkeywords}

\section{Introduction}

The rise of ultra-massive multiple-input multiple-output (UM-MIMO) systems, featuring thousands of antenna elements (AEs)~\cite{Wang2024Tutorial,lu2023tutorial}, is shaping future wireless communications. Driven by bandwidth demands, operating frequencies are shifting from sub-\unit[6]{GHz} to upper mid-band~\cite{Bjornson2024Enabling}, millimeter-wave, and terahertz (THz) bands~\cite{sarieddeen2021overview}. Such advances in spatial and spectral degrees of freedom are pivotal for meeting the increasing demands of next-generation networks.

The combination of UM array apertures and extremely short wavelengths fundamentally alters the electromagnetic propagation characteristics. In particular, far-field assumptions no longer hold universally, and near-field communication becomes inevitable even at moderate communication distances~\cite{Liu2024Near,tarboush2024cross}. 
Unlike the far-field planar wave model (PWM) that solely accounts for phase variations, the radiative near-field necessitates the spherical wave model (SWM) to accurately capture both amplitude and phase variations, which significantly increases signal processing complexity.

Extensive research has theoretically defined the near-field/far-field boundary using various criteria, including phase variation (e.g., the classical Rayleigh distance (RD) and its MIMO extensions~\cite{lu2023near}), array and beamforming gains (effective Rayleigh distance (ERD)~\cite{cui2024near} and Bj{\"o}rnson distance~\cite{bjornson2021primer}), channel gain and rank~\cite{li2025applicable,sun2025differentiate}, signal amplitude and phase variations~\cite{lu2021communicating}, channel capacity~\cite{jiang2005spherical}, and bandwidth-aware near-field criteria for wideband systems~\cite{deshpande2022wideband}, among others~\cite{bohagen2009spherical}. Nevertheless, evaluating these boundaries typically requires prior knowledge of parameters such as the communication distance, array aperture, and angle-of-arrival or departure. Unfortunately, in practice, such information is generally unavailable during the initial access stage and is instead estimated after link establishment. Our prior works~\cite{tarboush2024cross,tarboush2024near} have demonstrated that identifying the appropriate propagation regime before performing channel estimation or beamforming can significantly improve estimation accuracy while reducing computational complexity. However, these solutions rely on heuristic rules to determine the field boundaries and degrade in low signal-to-noise ratio (SNR) regimes, which are common during initial access in practical systems before configuring the beamformer and combiner. Moreover, these approaches lack robustness across diverse and dynamic environments.

Motivated by these challenges, this work proposes a fully unsupervised learning framework to distinguish near-field and far-field propagation regions based solely on received signal measurements. The proposed method does not rely on channel state information, user location, or labeled training data. Specifically, within an array-of-subarrays (AoSA) architecture, pairwise power differences between received signals across subarrays are computed and used as physics-informed features. These features are then processed using the OPTICS clustering algorithm to infer the underlying propagation regime.

\section{System Model and Problem Formulation}

\begin{figure}[htb!]
  \centering
  \includegraphics[width = 0.9\linewidth]{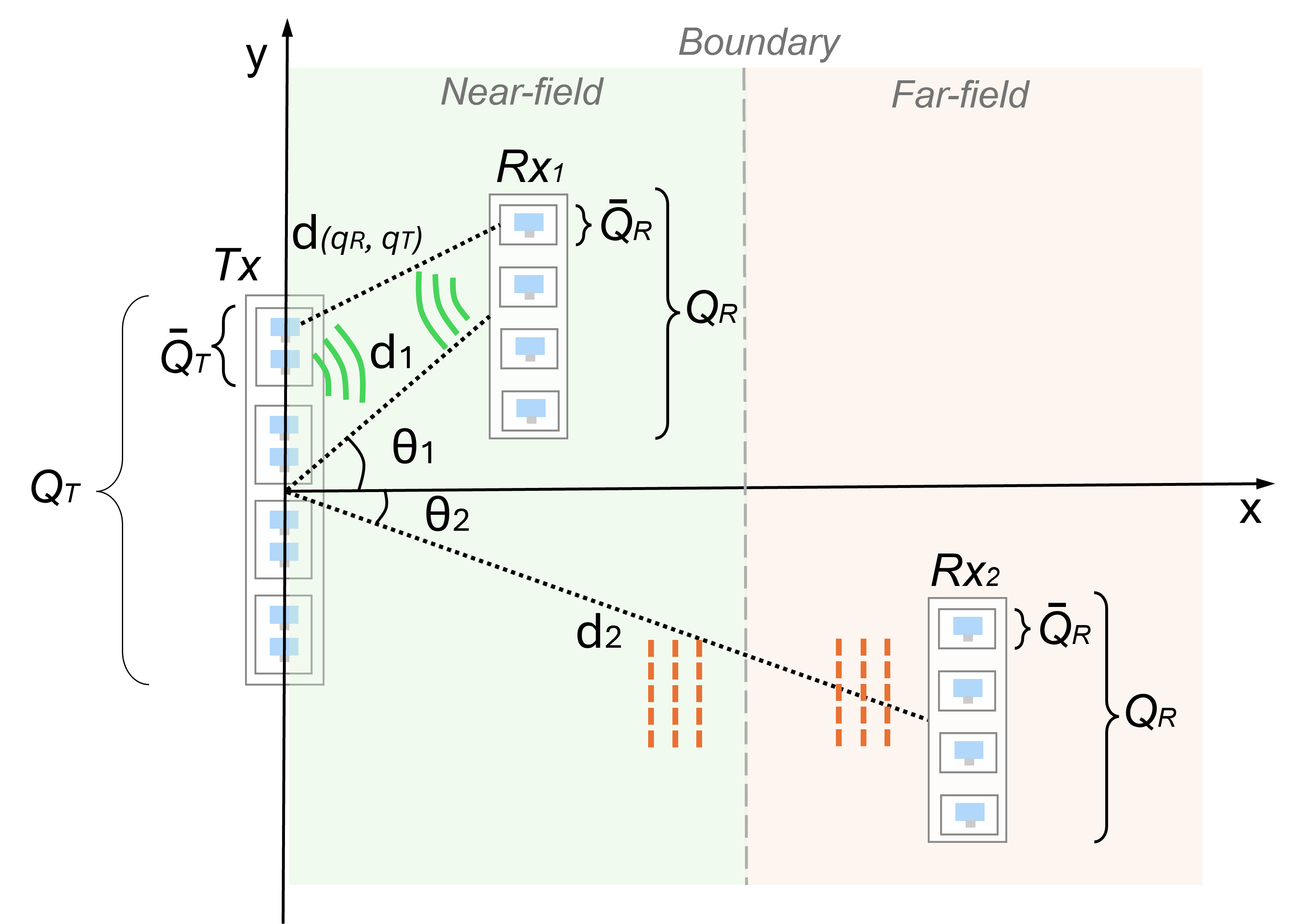}
  \caption{Illustration of near- versus far-field region identification problem, where the base station and multiple users employ an UM-MIMO AoSA architecture.}
  \label{fig:sys_diag}
\end{figure}

We consider a downlink narrow-band multi-user single-carrier system with bandwidth $B$ and center frequency $f_c$. We further assume an AoSA architecture of $Q_{\txidx}$-transmit and $Q_{\rxidx}$-receive subarrays (SAs), where each transmitter and receiver SA adopt a uniform linear array (ULA) of $\bar{Q}_{\txidx}$ and $\bar{Q}_{\rxidx}$ AEs, respectively, resulting in a $Q_{\rxidx}\bar{Q}_{\rxidx}\times Q_{\txidx}\bar{Q}_\txidx$ UM-MIMO configuration, as illustrated in Fig.~\ref{fig:sys_diag}. Each SA is connected to only one radio frequency (RF)-chain, respecting power and hardware constraints, and every AE is connected to an analog finite-resolution phase-shifter. Specifically, for each $\nth{q_\txidx}$ SA, the beamforming vector has a constant-magnitude and a variable-phase shift obtained from the set $\{0,\frac{2\pi}{2^{Q_\txidx^\quantidx}},\dots,\frac{2\pi(2^{Q_\txidx^\quantidx}-1)}{2^{Q_\txidx^\quantidx}}\}$, for a $Q_\txidx^\quantidx$-bit phase-shifter. At the receiver, the RF combiner follows the same assumptions as the transmitter.

The accurate channel model assumes SWM at any communication distance. The corresponding frequency-selective time-invariant channel, between $\nth{q_\txidx}$ transmit (Tx) SA and $\nth{q_\rxidx}$ receive (Rx) SA, for an UM-MIMO channel is constructed by collecting all channels gains, $h_{(\bar{q}_\rxidx,\bar{q}_\txidx)}^\SWMsupsc$, between the $\nth{\bar{q}_\txidx}$ transmit and $\nth{\bar{q}_\rxidx}$ receive AEs~\cite{tarboush2021teramimo}
\begin{equation}
\label{eq:overall_HSWM_smallh}
    \mbf{H}_{q_\rxidx,q_\txidx}^\SWMsupsc=\begin{bmatrix}
    h_{(1,1)}^\SWMsupsc& \cdots & h_{(1,\bar{Q}_\txidx)}^\SWMsupsc\\
    \vdots & h_{(\bar{q}_\rxidx,\bar{q}_\txidx)}^\SWMsupsc & \vdots\\
     h_{(\bar{Q}_\rxidx,1)}^\SWMsupsc & \cdots & h_{(\bar{Q}_\rxidx,\bar{Q}_\txidx)}^\SWMsupsc\\
    \end{bmatrix},
\end{equation}
\begin{equation}  
    h_{(\bar{q}_\rxidx,\bar{q}_\txidx)}^\SWMsupsc=\sqrt{\frac{1}{L}}\sum_{\ell=1}^{L}\alpha^\ell(f_c,d^\ell_{(\bar{q}_\rxidx,\bar{q}_\txidx)})e^{-j\frac{2\pi}{\lambda_c}d^\ell_{(\bar{q}_\rxidx,\bar{q}_\txidx)}},
    \label{eq:H_SWM_allcomp}
\end{equation}
where $L$ denotes the number of paths, $\alpha^\ell$ is the $\nth{\ell}$ complex path gain, $\lambda_c=\frac{c_0}{f_c}$ is the wavelength, $c_0$ is the speed of light, and $d^\ell_{(\bar{q}_\rxidx,\bar{q}_\txidx)}$ is the length of the $\nth{\ell}$ propagation path between the $\nth{\bar{q}_\txidx}$ transmit and $\nth{\bar{q}_\rxidx}$ receive AEs. The path gains vary across AEs--even for line-of-sight--as path loss includes spreading and molecular absorption losses, both of which vary with carrier frequencies and communication distances (see~\cite{tarboush2024cross,tarboush2021teramimo} for further details and expressions of the channel model).

The PWM approximates the SWM when the communication distance between the transmitter and receiver greatly exceeds the array aperture. Thus, the distance $d^\ell_{(\bar{q}_\rxidx,\bar{q}_\txidx)}$ in~\eqref{eq:H_SWM_allcomp} would be approximated by far-field assumptions~\cite{lu2023near}, where the channel gains of two AEs in the far-field differ only in phase, dependent on inter-AE spacing and the angle-of-departure (AoD), $\theta$ (or angle-of-arrival (AoA), $\phi$)~\cite{tarboush2024cross}. Consequently, the channel response is given by
\begin{equation}
    \label{eq:ch_pwm_frequencydomain_vectnot}
    \mbf{H}_{q_\rxidx,q_\txidx}^\PWMsupsc \!=\!\sqrt{\frac{\bar{Q}_\rxidx \bar{Q}_\txidx}{L}}\sum_{\ell=1}^{L}\alpha^\ell(f_c,d_\FFsupsc^\ell)e^{-j\frac{2\pi}{\lambda_c}d^\ell_\FFsupsc}\mbf{a}_\rxidx(\phi^\ell)\mbf{a}_\txidx^\Tpow(\theta^\ell),\!
\end{equation}
where $d_\FFsupsc$ is calculated following the far-field approximations, $\theta,\phi\in[-\pi/2,\pi/2]$, and the transmit/receive array response vector $\mbf{a}_\txidx(\cdot)\in\CC^{\bar{Q}_\txidx\times1}/\mbf{a}_\rxidx(\cdot)\in\CC^{\bar{Q}_\rxidx\times1}$ is 
\begin{equation}
    \label{eq:ULA_steering_vector}
    \mbf{a}(\theta)=\frac{1}{\sqrt{\bar{Q}}}\left[e^{j\frac{2\pi}{\lambda_{c}}\delta\sin(\theta)\mbf{\bar{q}}}\right],
\end{equation}
where $\mbf{\bar{q}}=\left[0,1,\cdots,\bar{Q}-1\right]^\Tpow$ and $\delta$ is the AE spacing.

From~\eqref{eq:overall_HSWM_smallh} and~\eqref{eq:ch_pwm_frequencydomain_vectnot}, it is clear that the unknown parameters differ across channel models. Typically, the suitable channel model, and consequently the channel estimation and beamforming strategies, for the near-field is SWM, while PWM has proven its accuracy in the far-field region. However, in practical scenarios, a mobile terminal transitions between the near-field and far-field regions of a system. This necessitates determining the user's region before initiating any signal processing techniques at the receiver.

\section{Proposed Machine Learning Framework}
\label{sec:prop_stra}
Although supervised learning offers classification capabilities, it is unsuitable here due to the inherent simplicity of distinguishing between near-field and far-field signals with known labels. The second fundamental problem is that any distance labels are biased to the used ground-truth boundary; however, each boundary applies a different approach/metric to distinguish these regions.

The proposed solution begins with a beam training stage to collect several measurements. After that, the received signal of each RF chain is utilized to compute a power variation metric, denoted as $\eta$, which constructs the feature matrix that we input into our unsupervised clustering algorithm. The intuition behind $\eta$ is to encapsulate the variability in received power across all Rx SAs. In the second stage, the ordering points to identify the clustering structure (OPTICS)~\cite{Ankerst1999Optics} clustering algorithm is adopted as the core unsupervised learning part.

During the beam training procedure, the transmitter and receiver employ precomputed random beamforming and combining weights stored in a beam codebook, obtained through uniformly distributed random sampling of angles. The transmit beamforming weights are sent from the reference Tx SA while simultaneously receiving pilots through random combining at all Rx SAs. In this phase, $\mbf{f}_{m_\txidx}$ and $\mbf{w}_{m_\rxidx}$ denote the beamforming and combining weight vectors, respectively, where $M_\txidx$ is the number of transmit beamforming measurements, while $M_\rxidx$ denotes the distinct combining measurements at the receiver. Consequently, the received signal, after a total of $M_\rxidx\times M_\txidx$ training beams for the $\nth{q_\txidx}$ Tx SA, can be expressed as~\cite{tarboush2024cross}
\begin{equation}
    \mbf{Y}_{q_\rxidx,q_\txidx}=\sqrt{P}\mbf{W}^\Hpow\mbf{H}_{q_\rxidx,q_\txidx}^\SWMsupsc\mbf{F}+\mbf{W}^\Hpow\mbf{N},
    \label{eq:allmeasure_trainingphase_rxsig_subck}
\end{equation}
where $P$ is the total power per transmission during the training phase, $\mbf{W}={\left[{\mbf{w}_1,\mbf{w}_2,\cdots,\mbf{w}_{M_\rxidx}}\right]}^\Tpow$ is the $\bar{Q}_\rxidx\times M_\rxidx$ combining matrix, $\mbf{F}={\left[{\mbf{f}_1,\mbf{f}_2,\cdots,\mbf{f}_{M_\txidx}}\right]}^\Tpow$ is the $\bar{Q}_\txidx\times M_\txidx$ beamforming matrix, and $\mbf{N}$ is an $M_\rxidx\times M_\txidx$ noise matrix comprising noise vectors, $\mbf{n}\sim\mathcal{CN}\left(\mbf{0},\sigma^2_n\mbf{I}\right)$, where $\sigma^2_n$ is the noise power.

Although determining the field boundaries seems complicated, one can utilize the physical characteristics of the near and far fields. Specifically, we can expect minimal signal variation across SAs in the far-field region, whereas significant variations occur in the near-field. Hence, we define the normalized vector as $\boldsymbol{\chi}^{q_r,1}=\frac{\abs{\vect{\mbf{Y}_{q_\rxidx,1}}}}{\norm{\vect{\mbf{Y}_{q_\rxidx,1}}}}$ following the vectorized version of the received signal in~\eqref{eq:allmeasure_trainingphase_rxsig_subck}. The selection metric $\eta$ captures the maximum variation in the received power across all Rx SAs and is defined as
\begin{equation}
    \label{eq:eta_metric}
    \begin{split}
    \eta&=\max{\{[\eta_{1,2},\!\cdots\!,\eta_{r,c},\!\cdots\!,\eta_{Q_\rxidx-1,Q_\rxidx}]\}}\\&
    \eta_{r,c} = (\boldsymbol{\chi}^{q_r,1}-\boldsymbol{\chi}^{q_c,1})^\Hpow(\boldsymbol{\chi}^{q_r,1}-\boldsymbol{\chi}^{q_c,1}),
    \end{split}
\end{equation}
where $r=\{1,\cdots,Q_\rxidx-1\}$, $c=\{r+1,\cdots,Q_\rxidx\}$. The output (values of $\eta$) of such a stage is equivalent to a feature extraction procedure in a typical machine learning pipeline.

With the preprocessed features, the framework utilizes OPTICS clustering for unsupervised learning. Unlike conventional clustering methods such as K-means or hierarchical algorithms, which require the number of clusters to be known in advance and often assume spherical or evenly sized groups, OPTICS is designed to discover clusters of arbitrary shape and density. This makes it particularly suitable for the problem at hand, where the spatial transition between near-field and far-field regions is often gradual and non-uniform, strongly influenced by antenna geometry, user location, and the underlying propagation environment. OPTICS operates by scanning all input data points, the $\eta$ values across receiver positions, and evaluating how closely each point is connected to its neighbors. Instead of assigning points to clusters directly, it produces an ordered list of points that reflects how tightly or loosely each point is linked to surrounding points, enhancing its ability to handle multi-density clusters.

\section{Numerical Results}

To evaluate the performance of the proposed framework, we simulate a THz-band UM-MIMO system using the TeraMIMO channel simulator~\cite{tarboush2021teramimo}. The system parameters are summarized in Table~\ref{table:simulationpara}. We emphasize that the SWM channel model is used in all simulations as it is the most accurate model for all communication distances. Measurements are collected across a two-dimensional $30$m by $30$m grid, with a resolution of $\unit[1]{m}$. At each point, and for every measurement configuration, several channel and noise realizations are performed. We emphasize that neither the channel matrix nor the receiver's $x-y$ positions are directly available to our model. Within our framework, although OPTICS does not require a predefined number of clusters, a binary near-field/far-field decision is ultimately required. The implementation orders the OPTICS clusters according to their mean power-variation feature. We first rank the clusters according to their mean feature value, and the maximum cluster is assigned to the near field, because larger $\eta$ means stronger spatial power variation. All remaining clusters are assigned to the far field. Then noise points are assigned according to the nearest observed $\eta$. Then, such binary clustering is performed while still allowing OPTICS to naturally identify consistent, well-separated regions based on the input features. Note that we have also tried several clustering algorithms, such as K-means and Agglomerative clustering. However, the identification was meaningless. Hence, only essential results are presented in this paper to convey the main findings.

\begin{figure*}[htb]%
 \centering
 \subfloat[$\mathrm{SNR} = 0 \; \mathrm{dB},M_\txidx=\bar{Q}_\txidx/2$]{\label{fig:sima} \includegraphics[width=0.31\textwidth,height=0.23\textwidth]{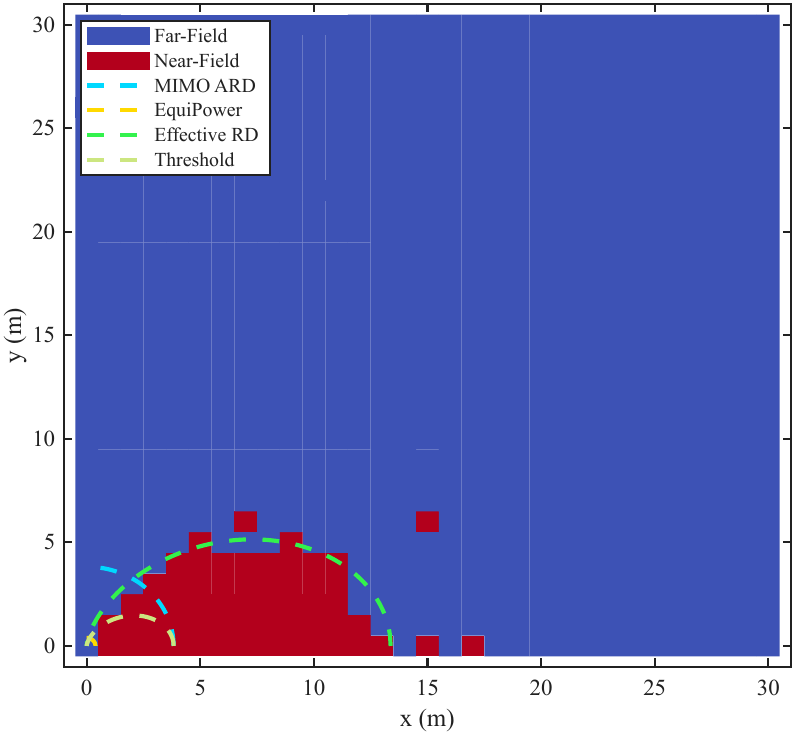}}
 \hfill
 \subfloat[$\mathrm{SNR} = 0 \; \mathrm{dB},M_\txidx=\bar{Q}_\txidx/4$]{\label{fig:simb} \includegraphics[width=0.31\textwidth,height=0.23\textwidth]{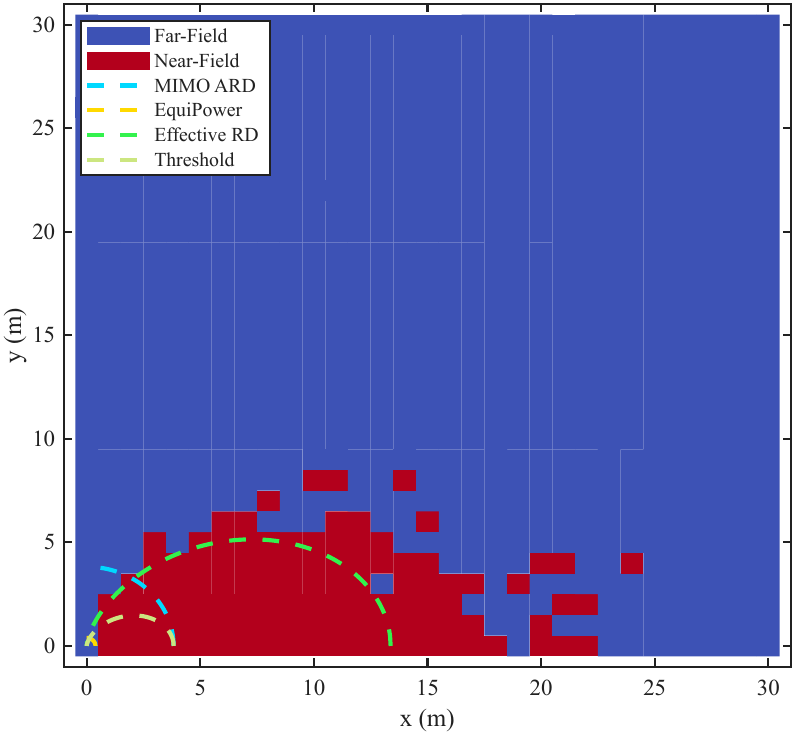}}
 \hfill
 \subfloat[$\mathrm{SNR} = 0 \; \mathrm{dB},M_\txidx=\bar{Q}_\txidx/8$]{\label{fig:simc} \includegraphics[width=0.31\textwidth,height=0.23\textwidth]{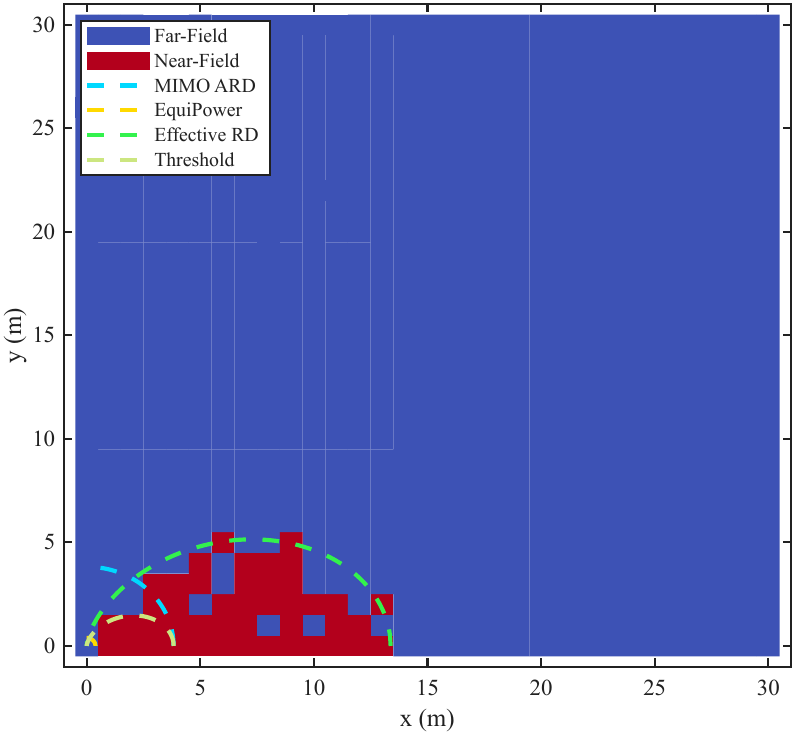}}
 \vspace{-3mm}
 \vfill
  \subfloat[$\mathrm{SNR} = -4 \; \mathrm{dB},M_\txidx=\bar{Q}_\txidx/2$]{\label{fig:simd} \includegraphics[width=0.31\textwidth,height=0.23\textwidth]{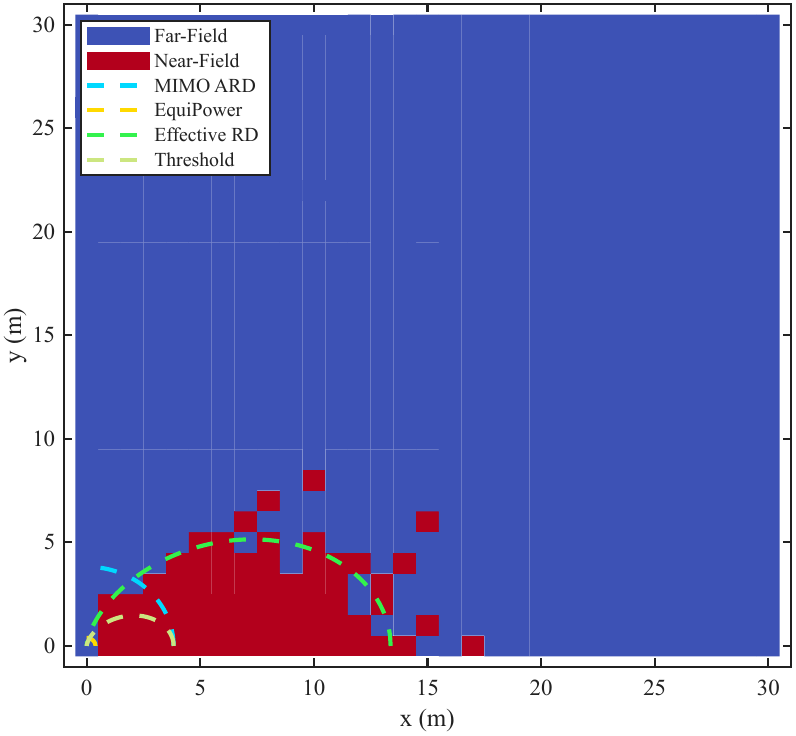}}
 \hfill
 \subfloat[$\mathrm{SNR} = -4 \; \mathrm{dB},M_\txidx=\bar{Q}_\txidx/4$]{\label{fig:sime} \includegraphics[width=0.31\textwidth,height=0.23\textwidth]{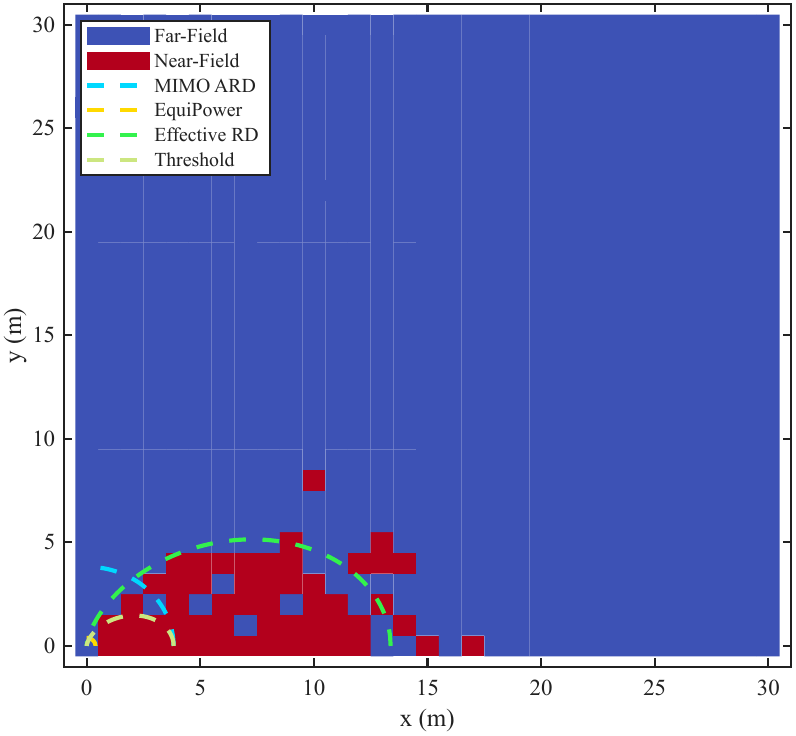}}
 \hfill
 \subfloat[$\mathrm{SNR} = -4 \; \mathrm{dB},M_\txidx=\bar{Q}_\txidx/8$]{\label{fig:simf} \includegraphics[width=0.31\textwidth,height=0.23\textwidth]{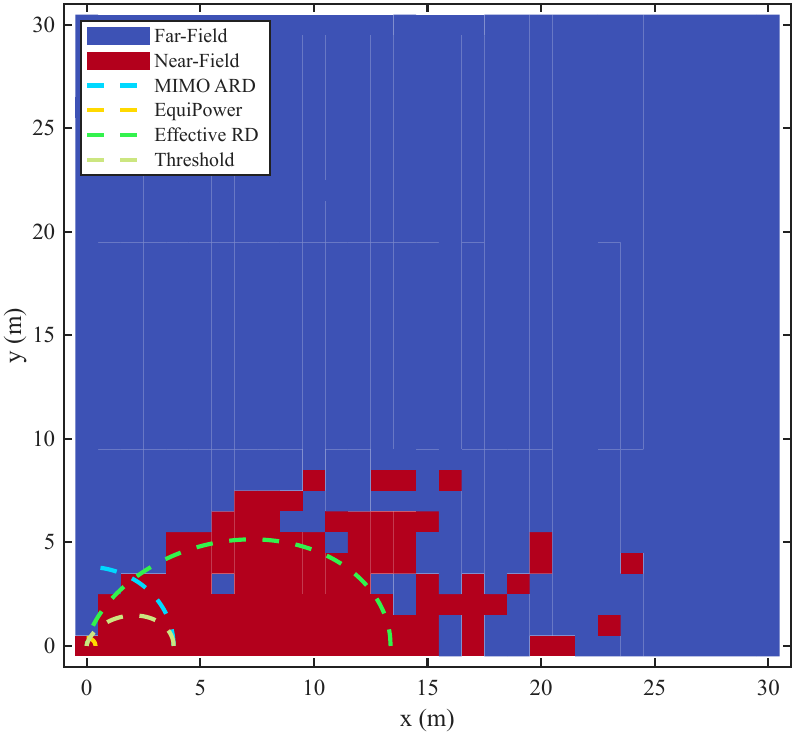}}
 \caption{Near-field (red) and far-field (blue) region identification under different SNR levels and beam training configurations.}%
 \vspace{-5mm}
 \label{fig:res}
\end{figure*}

While evaluating the numerical results in Fig.~\ref{fig:res}, we also added some reference boundaries, with the analytical expressions in~\cite[Table I]{sun2025differentiate}, defined as follows
\begin{itemize}
    \item MIMO-ARD~\cite{lu2023near}: The maximum phase error between the MIMO channel and the channel model by adopting the near-field array response vectors does not exceed $\frac{\pi}{8}$.
    \item Effective RD (ERD)~\cite{cui2024near}: The normalized beamforming gain under the far-field model is no less than $95\%$ of the beamforming gain computed using the near-field.
    \item Equi-power line~\cite{li2025applicable}: The normalized received power is approximately equal to one.
    \item Threshold distance~\cite{bohagen2009spherical}: The ratio between the largest eigenvalues of the SWM and PWM is equal to a predefined threshold.
\end{itemize}

\begin{table}
\footnotesize
\centering
\caption{Simulation parameters}
\begin{tabular} {|c || c|}
 \hline
 Parameters & Values\\ [0.5ex] 
 \hline
 \hline
 Operating frequency $f_c$ & $\unit[0.3]{THz}$ \\
 System bandwidth $B$ & $\unit[1]{GHz}$\\
 Tx/Rx SAs $Q_\txidx/Q_\rxidx$ & $1/4$ \\
 Tx/RX AEs $\bar{Q}_\txidx/\bar{Q}_\rxidx$ & $256/16$\\
 AEs spacing $\delta_\txidx=\delta_\rxidx$&$\{\frac{\lambda_c}{2}\}$\\
  SAs spacing $\{\Delta_\txidx,\Delta_\rxidx\}$&$\{\bar{Q}_\txidx \delta_\txidx,\bar{Q}_\rxidx \delta_\rxidx\}$\\
  Number of training beams $M_\txidx/M_\rxidx$ & $ \frac{\bar{Q}_\txidx}{n}/\bar{Q}_\rxidx, n \in \{2,4,8\}$\\
  Rx SNR & \{-4, 0, 4, 8\} (dB)\\
 \hline
\end{tabular}
\label{table:simulationpara}
\end{table}

Fig.~\ref{fig:res} shows that the proposed solution consistently identifies well-separated near-field (red) and far-field (blue) regions. Even with different training measurement ratios, the classification remains robust, with only minimal loss of accuracy at the field transition boundaries. However, at SNR$=\unit[-4]{dB}$, some degradation is observed in the form of scattered false points, especially at the lowest training rate. We also omitted the high SNR plots since this is not a practical scenario. Furthermore, surprisingly, the clustering results align remarkably well with the ERD boundary~\cite{cui2024near}, which makes sense since we used far-field random beam training and measured the power differences between the received signals. Again, in practice, the communication distance is estimated after the channel estimation and localization procedures. This strong agreement with ERD is particularly interesting because our framework requires no knowledge of user position, propagation distance, or array geometry. Regarding the computational complexity, although the raw received measurements can be high-dimensional, after feature extraction, the OPTICS inputs are a set of scalar $\eta$ observations (one per position). Moreover, the feature extraction scales with the number of receiver SAs rather than the total number of AEs. Then, the proposed approach provides a scalable solution with a small computational load.

\section{Conclusion}

This paper presented a fully unsupervised learning framework for near-field and far-field identification in systems equipped with UM-MIMO arrays. We proposed an unsupervised learning framework that effectively categorizes near-field and far-field signals based on their inherent characteristics and determines whether the current user is in the near or far region, even at low received SNR levels.  By utilizing the received signal variation as a feature and applying OPTICS clustering, we achieved robust classification without requiring user location, channel ground truth, or supervised training.

\bibliographystyle{IEEEtran}
\bibliography{abbrev,bibliography}

\end{document}